\pdfoutput=1

\documentclass[11pt]{article}

\usepackage[]{EMNLP2023}

\usepackage{times}
\usepackage{latexsym}
\usepackage{graphicx}
\usepackage{booktabs}
\usepackage{tabularx}
\usepackage{url}
\usepackage[T1]{fontenc}
\usepackage[utf8]{inputenc}
\usepackage{microtype}
\usepackage{inconsolata}
\usepackage{fancyvrb}

\title{Addressing the Blind Spots in Spoken Language Processing}

\author{
    Amit Moryossef \\
  Bar-Ilan University \\
  University of Zürich \\
  \texttt{amitmoryossef@gmail.com}}

\begin{document}
\maketitle

\begin{abstract}
This paper explores the critical but often overlooked role of non-verbal cues, including co-speech gestures and facial expressions, in human communication and their implications for Natural Language Processing (NLP). We argue that understanding human communication requires a more holistic approach that goes beyond textual or spoken words to include non-verbal elements. Borrowing from advances in sign language processing, we propose the development of universal automatic gesture segmentation and transcription models to transcribe these non-verbal cues into textual form. Such a methodology aims to bridge the blind spots in spoken language understanding, enhancing the scope and applicability of NLP models. Through motivating examples, we demonstrate the limitations of relying solely on text-based models. We propose a computationally efficient and flexible approach for incorporating non-verbal cues, which can seamlessly integrate with existing NLP pipelines. We conclude by calling upon the research community to contribute to the development of universal transcription methods and to validate their effectiveness in capturing the complexities of real-world, multi-modal interactions.
\end{abstract}

\section{Introduction}

Human speech is typically accompanied by a dynamic combination of co-speech gestures and facial expressions, together forming an integral part of human communication. These non-verbal cues, far from being random or merely accessory, provide additional layers of meaning, clarify intention, emphasize points, regulate conversation flow, and facilitate emotional connection. They enrich our interactions and help convey complex or nuanced information that words alone might not capture. 

Co-speech gestures refer to the hand and body movements accompanying spoken discourse; they supplement verbal communication by offering additional information, such as object size or shape; they emphasize and make abstract concepts tangible, like gesturing upwards to signify an increase; they control the conversation flow, signaling a speaker's intent, inviting listener interaction, or showing that the speaker is in thought or pause; and lastly, they compensate for the limitations of spoken language, especially in high-stakes or noisy environments, by providing an alternative mode of conveying complex or nuanced information.

Facial expressions during speech significantly contribute to communication by indicating the speaker's emotions, and providing insight into their feelings about the topic; they can emphasize certain aspects of the discourse, with actions like raised eyebrows signifying surprise or importance; they offer social cues, with expressions like a smile suggesting friendliness or a serious look indicating sincerity; they help clarify verbal meaning, especially in ambiguous situations, for example, a confused expression might denote misunderstanding; finally, they enhance interpersonal connection by helping to build rapport, expressing empathy, and conveying cues of understanding and engagement; altogether, facial expressions, like gestures, add complexity and depth to verbal communications.

The field of Natural Language Processing (NLP) has become highly effective in understanding language directly from text. However, understanding speech, with its imperfect and noisy signals, remains a more complex challenge. Text-based language models have proven highly scalable, thanks largely to the compressible nature of text and its abundant availability in semi-anonymous forms. Yet, these models fundamentally ignore the rich layers of meaning added by non-verbal cues, a significant aspect of human communication. This means that while we have become adept at parsing text, we are missing out on the nuanced interplay of speech and gesture that characterizes in-person communication. Despite some promising work in generating co-speech gestures from audio \cite{ginosar2019gestures, Bhattacharya2021Speech2AffectiveGesturesSC,liu2022audio}, these gestures are often treated as accessory to speech rather than integral components, and thus, they do not always contribute the correct or intended information. As such, an understanding and integration of non-verbal cues remain an important frontier for further exploration in NLP.

Spoken language understanding, we propose, can benefit immensely from the advances in sign language processing. We advocate for the implementation of universal automatic gesture segmentation and transcription models that can transcribe co-speech gestures into textual input. This could be a pioneering step towards integrating the richness of non-verbal cues directly into the NLP models. By including transcribed gestures, the models would bridge the blind spots in spoken language understanding.
This is a bidirectional process; Just as spoken language models can learn from sign language processing, the insights from the transcription of spoken language gestures can also inform and enhance sign language processing, due to iconicity, and metaphors. Ultimately, this holistic approach would result in a more nuanced and comprehensive understanding of human communication, bringing us closer to the complexities and richness of real-world, multi-modal interactions.

\section{Stereotypical Language Variation}

Non-verbal forms of communication are subject to significant cultural variability, shaped by a complex interplay of historical, societal, and cultural factors. 

In Mediterranean cultures, non-verbal communication is prevalent and vibrant. People in this region often use expressive gestures and maintain close personal space when communicating.
Italian, for instance, is renowned for its extensive use of gestures. Italians often use their hands and bodies expressively to illustrate their points or emotions, and there is a broad range of specific gestures that carry particular meanings, often comprehensible even without accompanying speech.

In contrast, Japanese communication tends to incorporate fewer and more subtle non-verbal cues. A bow, a nod, or a slight tilt of the head can convey a myriad of meanings depending on the context, demonstrating respect, agreement, or understanding. Meanwhile, in Nordic cultures, such as Swedish or Finnish, non-verbal cues are typically used sparingly. The communication style tends to be direct and understated, with less emphasis on gestures and more focus on verbal content.

Overall, these stereotypical examples highlight the diverse ways in which languages around the world incorporate non-verbal cues into communication. This diversity emphasizes the importance of cultural understanding and sensitivity in interpreting and engaging in cross-cultural communication research, and data collection and annotation.

\section{Motivating Examples}

Non-verbal cues can act to affirm and reinforce the spoken words, thereby strengthening the communicated message. They can also undermine the verbal message, creating a contradiction between what is being said and the speaker's true intent or feelings. For NLP research to understand speech, it can not rely solely on audio (or textual transcription) to understand the intent of the speaker.

For example, saying `Perfect' while making a circle with the thumb and index finger often emphasizes approval and satisfaction. Similarly, nodding while saying `Yes' reinforces affirmation, underscoring the speaker's understanding or agreement.
On the other hand, saying `OK' while rolling one's eyes, can suggest that the speaker doesn't find the situation truly satisfactory, despite the verbal agreement. Similarly, stating ``I'm not mad'' while frowning or clenching fists suggests that the speaker is indeed upset, contradicting their verbal assertion.

\subsection{Machine Translation}
While existing in many other languages, Italian stereotypically gives us many examples of gestures conveying meaning, where the verbal part is often dropped altogether, making it even more similar to signed languages.

Table \ref{table:italian} showcases a toy example of a conversation between two Italians using only gestures, without speech. It is transcribed using SignWriting \citep{writing:sutton1990lessons} to demonstrate that anonymous non-verbal transcription can be done in a low-bandwidth manner and that it can be reproduced and understood by people trained at reading SignWriting.

\begin{table*}[]
\centering
\begin{tabular}{*{4}{m{0.22\textwidth}}}
\toprule
\includegraphics[width=\linewidth]{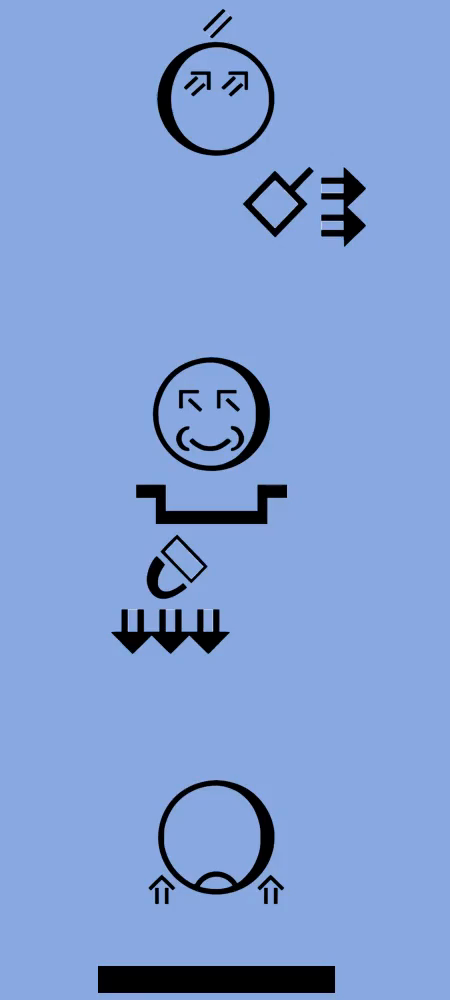} & 
\includegraphics[width=\linewidth]{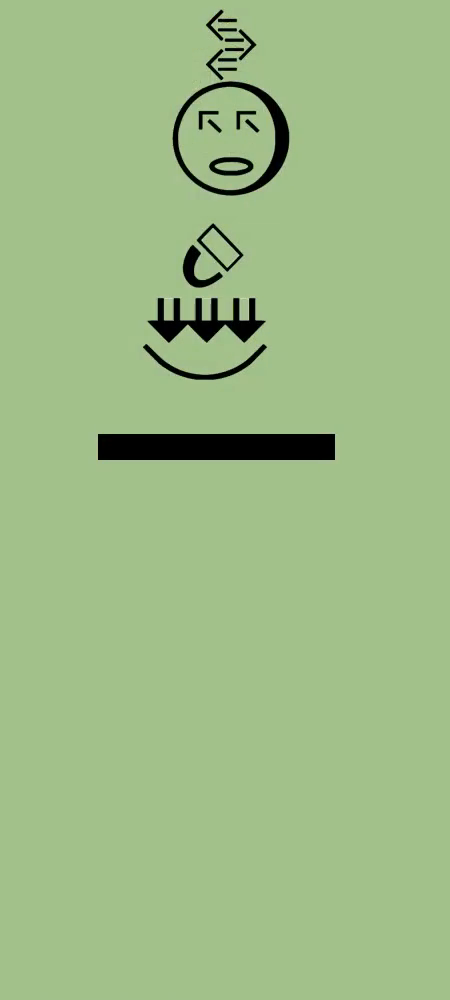} & 
\includegraphics[width=\linewidth]{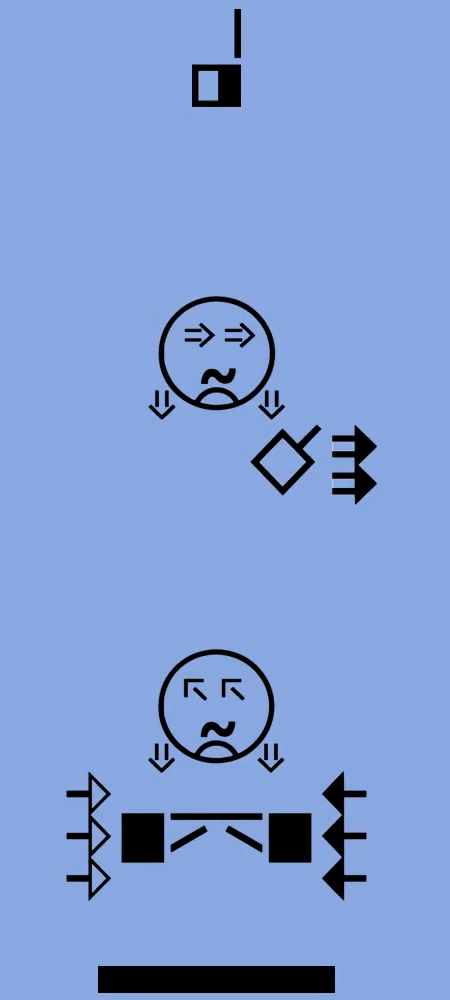} & 
\includegraphics[width=\linewidth]{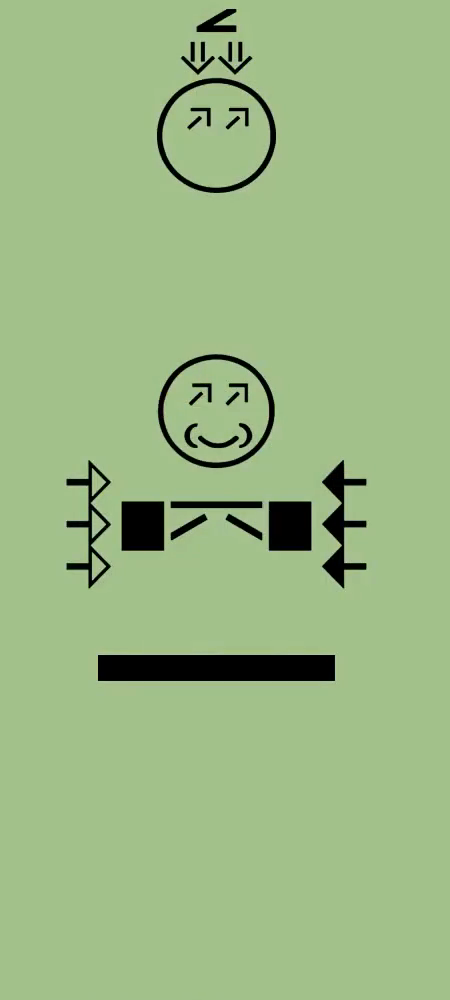} \\
\midrule
wth with these two?? & I know! What the hell? & Did you hear they're together?? & Oh yeah. They're definitely together! \\
\midrule

\includegraphics[width=\linewidth]{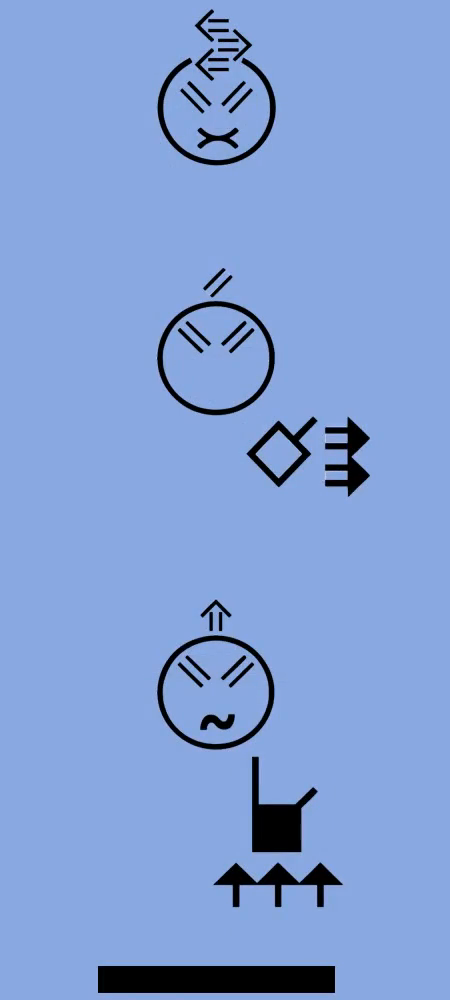} & 
\includegraphics[width=\linewidth]{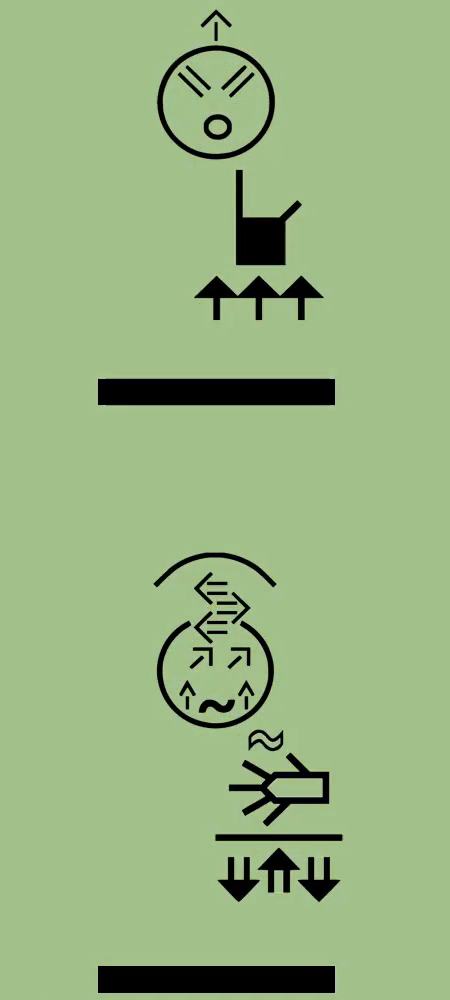} & 
\includegraphics[width=\linewidth]{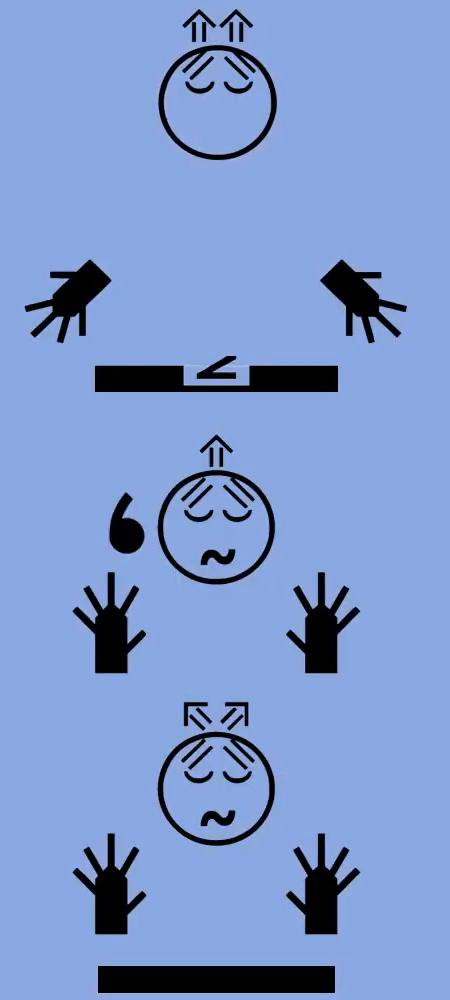} & 
\includegraphics[width=\linewidth]{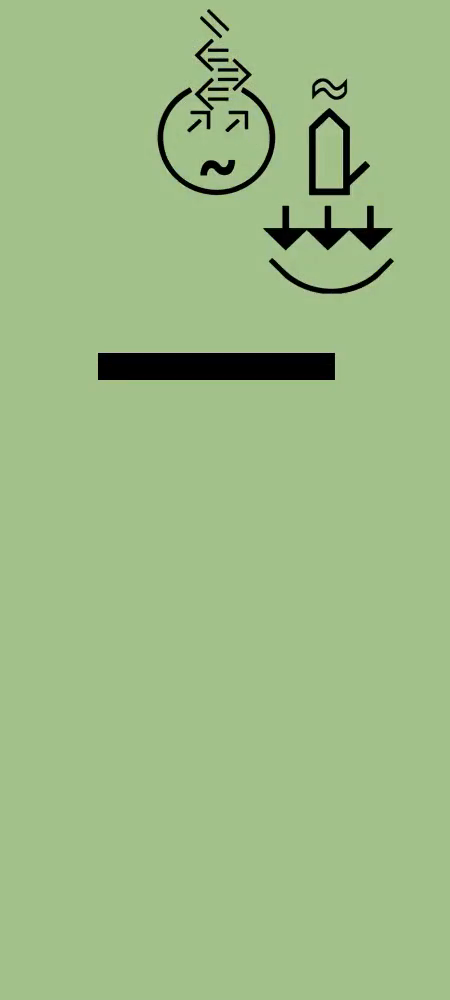} \\
\midrule
I shouldn't be saying this but... she's cheating on him & She's cheating on him?? Woahhh & Yeah! But it's none of my business & I can't believe it \\
\bottomrule
\end{tabular}

\caption{``How to gossip in Italian'' by the Pasinis, transcribed in Sutton SignWriting by Sutthikhun Phaengphongsai \url{https://www.youtube.com/watch?v=7V-GniCQFkE}, demonstrating a conversation between two Italians using only gestures, without speech.}
\label{table:italian}

\end{table*}

\subsection{Sentiment Analysis}

To demonstrate the limitations of text-based sentiment analysis, consider the following hypothetical dialogue between a couple, where the man is utilizing passive-aggressive communication. In each turn, we also present the sentiment score as predicted by the Google Cloud Natural Language API Demo\footnote{\url{https://cloud.google.com/natural-language}}, where scores range between $[-1, 1]$.

\begin{Verbatim}[commandchars=\\\{\}]
  Woman: How is it going? (\textcolor{yellow}{0})
  Man:   I am fine. (\textcolor{green}{0.74}) 
         \textcolor{gray}{[crosses his arms]}

  Woman: Did you enjoy dinner? (\textcolor{green}{0.55})
  Man:   It was fine. (\textcolor{green}{0.92}) 
         \textcolor{gray}{[avoids eye contact,} 
         \textcolor{gray}{lips pressed tightly]}

  Woman: Is something wrong? 
         You seem distant. (\textcolor{red}{-0.78})
  Man:   No, nothing's wrong. (\textcolor{green}{0.53}) 
         \textcolor{gray}{[shakes his head slightly,}
         \textcolor{gray}{exhales loudly]}

  Woman: Are you sure? (\textcolor{yellow}{0})
  Man:   I said I'm fine. (\textcolor{yellow}{0}) 
         \textcolor{gray}{[rolls eyes, turns away]}
\end{Verbatim}

While all the man's responses register as neutral to positive, his body language—avoiding eye contact, pressing his lips tightly together, shaking his head, exhaling loudly, rolling his eyes, and turning away—signals that he may actually be upset, frustrated, or disengaged.
By neglecting body language and other contextual clues, current models miss out on a significant layer of human communication, particularly in emotionally charged or complex dialogues. Such a holistic approach could provide a more nuanced and accurate understanding of the emotional context and underlying issues, thus enriching machine-human interactions.

\section{Methodology}

Machine learning techniques that focus solely on text have gained predominance due to several key factors: the abundance of readily available text data, the potential for semi-anonymous data collection and processing, the high bandwidth-to-overhead ratio as a word consumes only a few bytes compared to kilobytes or more for a second of speech or video, and the ease with which text can be viewed, edited, and corrected.

Previous efforts have attempted to include other modalities like images \citep{Razavi2019GeneratingDH}, videos \citep{Yan2021VideoGPTVG}, or audio through the use of techniques like VQ-VAEs \citep{Oord2017NeuralDR}. However, these approaches often significantly increase the context size, are not transferable across different systems, and generally require the original signal (like a video) to be sent for processing. In contrast, our proposal offers a more flexible, universal, and efficient way to incorporate non-verbal cues directly as text.

\subsection{Proposal}

We propose adopting a universal transcription system for body language, much like the written system used for spoken languages. This system would transcribe gestures, facial expressions, and other non-verbal cues into textual form. The advantages of this approach are numerous:

\paragraph{Flexibility in Transcription} Different programs can decide on their own transcription methods, taking into account local variations and context.
\paragraph{Computational Efficiency} Text-based methods require significantly lower computational resources compared to image or video processing. (Notoriously, GPT-4 was released without image upload support, since inference on a single image takes upwards of 20 seconds)
\paragraph{Compatibility with Existing Models} As the body language would be transcribed into discrete tokens, it can fit seamlessly into existing large language models without any modification.
\paragraph{Anonymity} Transcription acts as a form of biometric anonymization, removing the need to share actual video or images.
\paragraph{Explainability} The textual transcription provides a more transparent input, making the language modeling process more understandable.
\paragraph{Seamless Integration} The proposed methodology does not require any significant changes to existing NLP pipelines. It simply acts as an additional layer of data for better understanding and disambiguation. You can include it, or not.

\subsection{Implementation}

To successfully integrate non-verbal cues when processing spoken language, we advocate the following steps:

\begin{enumerate}
    \item Capture both video and audio during speech.
    \item Use sign language segmentation models to identify boundaries of individual gestures.
    \item Transcribe these gestures into a textual notation system like SignWriting.
    \item Use speech-to-text models to transcribe the spoken language, identifying the boundaries where each word is expressed.
    \item If word boundaries are not directly accessible, a re-alignment model can be used to approximate these boundaries.
    \item Combine both speech and gesture transcriptions into a single text string, where gestures can be used to provide additional context to the spoken words.
\end{enumerate}

This approach can be thought of as analogous to incorporating additional context, such as gender, into machine translation \citep{moryossef-etal-2019-filling}. By training on a large dataset that includes unmarked sentences, the model may develop certain biases. Introducing a smaller dataset with contextual information can help the model learn correlations between language and specific contexts. During inference, one has the option to either provide just the text for a more generalized output or include additional contextual tags for a more accurate and targeted output.

\section{Conclusions}

This paper underscores the fundamental role of non-verbal cues, such as co-speech gestures and facial expressions, in human communication. While strides have been made in the realm of Natural Language Processing for understanding textual content, a holistic approach that integrates the rich layers of non-verbal information is significantly lacking. This shortfall not only hampers the comprehension of spoken language but also limits our ability to construct nuanced, context-aware NLP models.

The key to advancing in this frontier may lie in borrowing techniques and insights from sign language processing. We advocate for the adaptation and implementation of universal automatic gesture segmentation and transcription models that can transcribe co-speech gestures into textual input. Such an approach would be a pivotal step in bridging the gap between text-based and real-world, spoken interactions, thereby enriching both the scope and applicability of NLP models.

When processing spoken language content, researchers should adopt a more holistic lens, one that goes beyond words and phrases to also consider non-verbal cues. Existing universal segmentation and transcription models used in sign language can serve as invaluable resources for this purpose, as they offer the ability to transcribe gestures directly as text. 

We call upon researchers in spoken language processing to contribute to the development of universal gesture transcription methods. Furthermore, we encourage the academic community to construct challenge sets specifically tailored to validate the utility of these transcription methods in capturing the complexities of non-verbal communication. These steps are not merely supplementary but are central to achieving a more comprehensive understanding of human communication in its full richness and complexity.

\bibliography{anthology,custom}
\bibliographystyle{acl_natbib}

\end{document}